# Plasma-Enhanced Germination in North Indian Wheat

Punit Kumar[a], Abhishek Kumar Singh[b] and Priti Saxena[c]
[a]Department of Physics, University of Lucknow, Lucknow – 226007, India
[b]Department of Physics, G L Bajaj Group of Institutions, Mathura - 281406, India
[c]Department of Zoology, D.A.V. Degree College, Lucknow – 226004, India

*Abstract*—The application of non-thermal plasma in agriculture has emerged as a sustainable and eco-friendly method to enhance seed vigor, germination, and crop productivity. This study investigates the effects of atmospheric pressure plasma treatment on five popular bread wheat varieties of North India, WH 1142, HI 1544, GW 366, GW 322, and GW 273. Direct dielectric barrier discharge (DBD) plasma exposure (1 – 5 min) and plasma activated water (PAW) irrigation (5 – 15 min activation) were tested. Results indicated significant improvements in seed wettability, germination index, root and shoot growth, spike length, and grain yield compared to controls. Among treatments, 3 min DBD exposure and 15 min PAW irrigation consistently produced the best results, with variety specific differences in vigor and yield. These findings demonstrate the potential of plasma seed treatment as a chemical free technology to enhance productivity in wheat, contributing to sustainable agriculture in India.

## I. INTRODUCTION

WHEAT (*Triticum aestivum* L.) is the second most widely cultivated cereal crop worldwide after rice. High yielding bread wheat varieties play a critical role in ensuring food security through both commercial cultivation and inclusion in national food supply systems. However, the productivity of wheat is frequently constrained by challenges such as declining seed vigor, nutrient stress, and the increasing impacts of climate variability. Poor seed quality adversely affects germination, stand establishment, and final yield. As seed vigor is a fundamental determinant of crop success, there is an urgent need for innovative and sustainable technologies that can enhance germination and early growth performance of wheat under diverse global growing conditions.

To overcome the limitations of poor seed vigor, several priming techniques have historically been employed. Chemical seed priming involves the use of growth regulators or salts to initiate early metabolic activity, but this often leaves behind toxic residues harmful to soil health and the environment [2]. Physical priming methods such as ultrasonic scratching, magnetic field exposure, ion beam irradiation, and electric field treatments have also been applied to stimulate germination and improve vigor [3-5]. While these approaches sometimes enhance enzymatic activity and improve germination, they also carry drawbacks: chemicals may contaminate soil and food chains, while certain physical techniques can damage seed cells, reduce viability, and lead to uneven results depending on treatment duration and conditions [6,7]. Further, such methods can be inconsistent and costly when applied at scale in large farming systems. This highlights the urgent requirement for a safe, reliable, and eco-friendly alternative for seed treatment in wheat production.

In this context, atmospheric pressure non-thermal plasma (APP) has emerged as a novel, chemical free technology for seed enhancement [8-10]. Plasma, often described as the fourth state of matter, consists of a mixture of electrons, ions, UV photons, and reactive oxygen and nitrogen species (RONS). Unlike thermal plasmas that operate at high temperatures, non-thermal plasmas are generated at room temperature and atmospheric pressure, making them safe for treating biological materials such as seeds. The nonequilibrium nature of these plasmas allows for a unique combination of physical and chemical processes that alter the seed surface and stimulate internal biochemistry without causing thermal damage. Previous studies have shown that atmospheric pressure plasma can modify seed surface roughness, reduce the contact angle of water droplets, and improve hydrophilicity, thereby enhancing water uptake and imbibition efficiency [11-14]. Reactive oxygen and nitrogen species created during plasma exposure, including ozone, hydroxyl radicals, nitrites, and nitrates, can penetrate seed coats, stimulating metabolic processes, breaking dormancy, and reducing microbial contamination on seed surfaces. Such processes collectively accelerate germination, promote uniform seedling growth, and improve plant vigor [15,16].

A substantial body of research has already demonstrated these benefits in different crops. Studies have shown that plasma exposure stimulates and accelerates biological processes in seeds, enhancing germination and early growth [17,18]. Plasma-induced reactive species (RONS) enhance water absorption [19], reduce dormancy [20], eliminate contaminants [21,22], improve germination percentage [23], accelerate seedling growth [24,25], and contribute to greater tolerance under drought and other stresses [26]. The positive effects of plasma treatment on seed germination and seedling vigor have been reported for multiple plant species [27,28], and surface modifications produced by plasma have been linked to better germination outcomes under varied environmental conditions [29]. It is also well recognized that seeds with high moisture content are vulnerable to freezing damage during imbibition [30], and enhancing hydrophilicity through plasma treatment allows seeds to absorb water more efficiently, improving early development [29]. However, the outcomes of plasma treatment depend strongly on the plasma generation method [31] and the working gas used [32], underlining the need for systematic evaluation of appropriate configurations and exposure times.

Plasma treatments on wheat seeds have demonstrated notable improvements in germination potential, growth rate, and yield performance, especially when using dielectric barrier discharge (DBD) plasmas and plasma-activated water (PAW). DBD plasma allows for direct treatment of seeds by exposing them to the discharge for short durations, while PAW is generated by treating water with plasma, leading to physico-chemical changes such as reduced pH, increased oxidation-reduction potential, and accumulation of nitrite and nitrate species. These changes enhance the fertilization capacity of irrigation water and provide an additional mechanism for improving crop performance. Previous work has shown that PAW can supply reactive nitrogen species necessary for growth, functioning as a natural fertilizer and promoting root elongation, spike length, and overall biomass accumulation [33,34].

The present research aims to systematically investigate the effects of atmospheric pressure plasma treatment on popular varieties such as WH 1142, HI 1544, GW 366, GW 322, and GW 273 using both direct DBD plasma exposure and indirect plasma-activated water application. Specifically, the study examines changes in seed surface characteristics, germination percentage, germination index, root and shoot growth, spike length, grain number per spike, root biomass, and thousand-grain weight. By combining these physiological and yield parameters, the study seeks to establish optimized plasma treatment protocols that can enhance seed vigor and productivity under North Indian field conditions. Ultimately, the significance of this research lies in its potential to provide farmers with a chemical free, environmentally sustainable method of improving wheat productivity, while contributing to broader goals of food security and climate resilient agriculture in India.

## II. MATERIALS AND METHODS

Certified seeds of five bread wheat (*Triticum aestivum L.*) varieties, WH 1142, HI 1544, GW 366, GW 322, and GW 273, were obtained from the authorized vendors. These varieties are among the most widely cultivated in the northern plains of India and represent a diverse genetic base, making them suitable for systematic evaluation of plasma treatments. To ensure experimental reliability, only uniform and healthy seeds were selected after manual cleaning and sorting. The study was designed to compare direct plasma exposure using dielectric barrier discharge (DBD) and indirect plasma treatment through plasma-activated water (PAW) with untreated controls.

For the direct plasma treatment, seeds were subjected to atmospheric pressure DBD plasma generated between two parallel circular brass electrodes of 30 mm diameter and 7 mm thickness, separated by a 1.8 mm polycarbonate dielectric barrier. The discharge was produced using a transformer operating at 50 Hz with a voltage range of 0–30 kV, with setup shown in Fig. 1. The resulting plasma had an electron excitation temperature of approximately 0.91 eV and an electron density of the order of $10^8$ cm$^{-3}$. Seeds were exposed for 2, 4, 6, 8 and 10 minutes, designated as treatments $T_1$, $T_2$, $T_3$, $T_4$ and $T_5$ respectively. The untreated seeds served as the control group ($T_0$). The exposure chamber was designed to avoid thermal damage, ensuring that the plasma remained non-thermal and safe for biological samples.

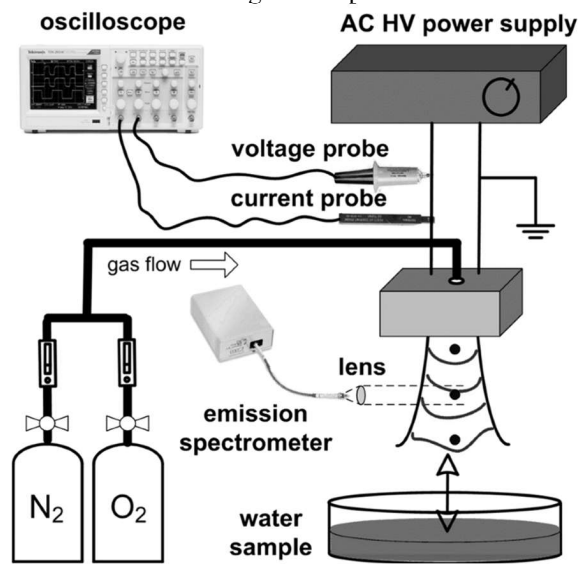

**Fig. 1 :** Schematic diagram of experimental arrangement

Plasma-activated water was prepared using a gliding arc discharge system. The discharge was produced between copper electrodes 50 mm long, separated by 1.5 mm at the inlet and 4 mm at the outlet, with a diameter of 1.5 mm. The discharge voltage ranged from 0–10 kV DC, and the resulting plasma had an electron excitation temperature of 1.15 eV and an electron density of $10^{10}$ cm$^{-3}$. Tap water (250 ml) was exposed to the discharge for durations ranging 5 to 15 minutes, generating PAW with progressively altered chemical properties. Zero-minute PAW was considered untreated water and used as control. The gap between the electrode tip and the water surface was maintained at 15 mm to optimize plasma – liquid interaction.

Physico - chemical properties of PAW were measured to characterize treatment effects. Parameters included pH, electrical conductivity (EC), total dissolved solids (TDS), and oxidation–reduction potential (ORP), all assessed with a 7-in-1 RCYAGO water quality tester. Concentrations of nitrate ($NO_3^-$) and nitrite ($NO_2^-$) ions were measured with commercial test strips. Previous studies suggest that the decrease in pH, increase in ORP, and accumulation of reactive nitrogen species in PAW play an important role in promoting germination and growth by acting as bioavailable nitrogen sources [35].

The experiments were conducted under controlled conditions with a temperature of 25–30°C and relative humidity between 70–80%, simulating typical North Indian Rabi season environments. Seeds were sown in square pots (25 × 25 cm) filled with a mixture of natural soil (50%) and vermicompost (50%). Each treatment consisted of three replicates with 20 seeds per pot. Sowing depth was maintained at approximately 4 cm. Irrigation was performed with either untreated water or PAW, depending on the treatment group. Standard greenhouse practices were followed, and no additional fertilizers or pesticides were applied to isolate the effect of plasma treatment.

Multiple parameters were recorded to assess treatment effects. Seed quality traits included measurements of contact angle and wettability, providing insights into seed surface hydrophilicity. Contact angle was measured using water droplet analysis, while wettability was calculated using the weight difference of seeds before and after soaking, following the formula, $Wettability = (m_1 - m_2)/m_0$ described in earlier studies [35]. Germination percentage (=seeds germinated in $1 day \times 100$/total number of seeds) and germination index (=no. of germination seeds in t days/germination days) were evaluated over an 7 day period, with germination potential expressed as the ratio of seeds germinated on the first day to total seeds [13], and germination index calculated as the sum of the number of germinated seeds divided by the corresponding day of observation.

Seedling traits were recorded at 7 and 14 days after sowing. Root and shoot lengths were measured manually with a scale, and growth rate was calculated using plant height differences across time intervals, expressed in cm per day [35]. These observations were used to assess early vigor and establish the physiological benefits of plasma treatment.

Yield attributes were evaluated at 120 days after sowing, corresponding to crop maturity. Plants were harvested, and spike length, root length, number of grains per spike, and thousand-grain weight were recorded. These yield-related traits were selected as they directly contribute to productivity and have been shown in previous plasma studies to respond positively to treatment.

All experimental data were analyzed statistically. One-way analysis of variance (ANOVA) was performed at a significance level of α = 0.05 to determine differences between treatments. Duncan's multiple range test was employed to compare treatment means and identify significant differences. Data analysis confirmed whether observed improvements in germination, seedling growth, and yield attributes could be attributed to plasma exposure rather than random variation.

By combining direct DBD exposure and indirect PAW irrigation under controlled greenhouse conditions, this study systematically examined the physiological and agronomic effects of plasma treatment on Indian wheat varieties. The methodological approach was designed to generate reproducible evidence on the suitability of plasma technology as a chemical free alternative for enhancing wheat seed vigor and productivity, while retaining continuity with global studies.

## III. RESULTS AND DISCUSSION

The present study systematically evaluated the influence of dielectric barrier discharge (DBD) plasma treatment and plasma-activated water (PAW) on certified seeds of five popular Indian bread wheat varieties, WH 1142, HI 1544, GW 366, GW 322, and GW 273 under controlled conditions. The observations focused on seed surface properties, germination performance, seedling vigor, and yield attributes, with data supported by statistical analysis, optical emission spectroscopy (OES), and water chemistry parameters. The findings confirm that plasma treatment induces favorable physico-chemical modifications in seeds and irrigation water, thereby enhancing wheat growth and productivity, and also highlight variety specific responses relevant to Indian agricultural contexts.

Plasma treatment significantly altered the seed surface, as measured by changes in water droplet contact angle. Representative images of untreated and 5 minute plasma treated seeds are shown in Fig. 2, while the variation in contact angle with increasing exposure time is presented in Fig. 3. The average contact angle decreased from 75.69 ± 1.12° in untreated seeds to 64.91 ± 1° after 5 minutes of DBD exposure, demonstrating enhanced hydrophilicity. This surface modification facilitates faster imbibition and higher water uptake, corroborating earlier studies [36]. Consistent trends were recorded across varieties, though the extent of change was most pronounced in HI 1544 and GW 366, suggesting that seed coat morphology may influence responsiveness.

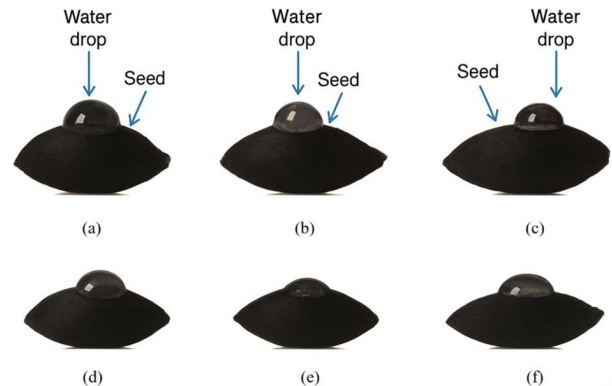

**Fig. 2 :** Water drop on surface of wheat seed (a) untreated, (b) WH 1142, (c) GW 322, (d) HI 1544, (e) GW 366, and (f) GW 273 respectively after 5 minutes of treatment.

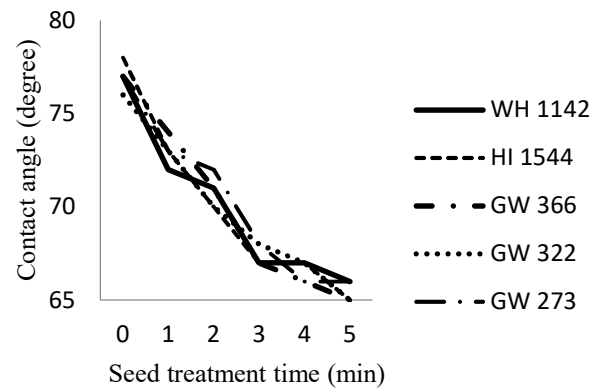

**Fig. 3 :** Variation of contact angle with time.

Wettability experiments reinforced these findings. As shown in Fig. 4, plasma-treated seeds absorbed significantly more water over time than controls. The wettability index increased steadily with longer plasma exposure, indicating structural modifications in the outer seed layers that allow water penetration. The improvement was greatest in GW 366, which exhibited mass increase during soaking. This aligns with the strong shoot vigor later recorded in this variety, indicating a linkage between surface modifications and early growth performance.

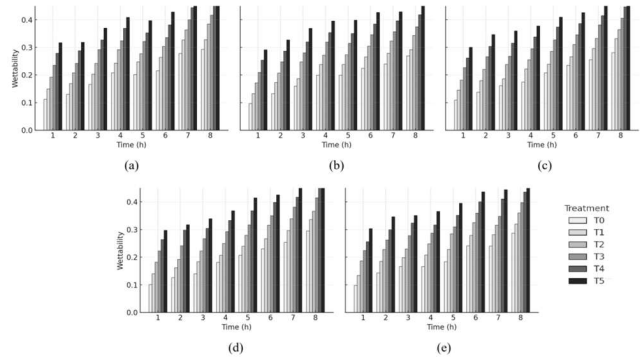

**Fig. 4 :** Wettability of (a) WH 1142, (b) HI 1544, (c) GW 366, (d) GW 322, and (e) GW 273 respectively, on soaking.

| Time (min) | pH | ORP (mV) | EC (µS/cm) | TDS (mg/L) | Nitrate (mg/L) | Nitrite (mg/L) |
|---|---|---|---|---|---|---|
| 0 | 7.0 ± 0.1 | 250 ± 5 | 500 ± 10 | 320 ± 5 | 5 ± 0.2 | 0.2 ± 0.01 |
| 1 | 6.8 ± 0.1 | 310 ± 6 | 520 ± 12 | 330 ± 4 | 15 ± 0.5 | 0.5 ± 0.02 |
| 3 | 6.5 ± 0.1 | 380 ± 8 | 540 ± 15 | 345 ± 6 | 25 ± 0.7 | 0.9 ± 0.03 |
| 5 | 3 ± 0.1 | 420 ± 7 | 560 ± 10 | 360 ± 5 | 30 ± 1.0 | 1.2 ± 0.05 |

*Values represent mean ± standard deviation (n = 3).*
**Table -1 :** Physico-chemical properties of Plasma-Activated Water (PAW) at different treatment times

The physico-chemical characterization of PAW is summarized in Table - 1. Treatment of water for 5, 10, and 15 minutes led to progressive decreases in pH from 6.2 to 5.2, while oxidation–reduction potential (ORP) increased from 250 mV to 420 mV. Electrical conductivity and total dissolved solids also rose with treatment time, reflecting the accumulation of ionic species. Nitrate ($NO_3^-$) concentrations increased from 12 mg/L in untreated water to 240 mg/L after 15 minutes, while nitrite ($NO_2^-$) levels rose from 0 to 39 mg/L. These results confirm the enrichment of water with reactive nitrogen species, consistent with earlier findings that such modifications enhance germination and growth [37]. For Indian wheat, PAW treatments were particularly effective for GW 366 and GW 322, where higher nitrate availability supported root elongation and grain filling.

Germination performance improved significantly in plasma-treated seeds across all varieties. Fig. 5 illustrates the germination potential, which increased markedly under 3–5 minute DBD treatments and with 15-minute PAW irrigation. WH 1142 and HI 1544 recorded the highest early germination percentages, establishing uniform stands within ten days of sowing. The germination index, shown in Fig. 7, was also significantly higher for plasma-treated seeds, with 3-minute DBD and 15-minute PAW treatments producing maximum values. These results indicate accelerated metabolic activation and improved seed vigor, corroborating earlier plasma studies on wheat [17,18].

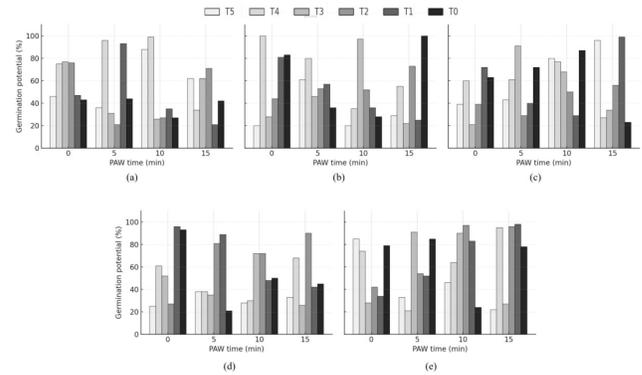

**Fig. 5 :** Variation of germination potential in % with plasma treatment time for (a) WH 1142, (b) HI 1544, (c) GW 366, (d) GW 322, and (e) GW 273 respectively.

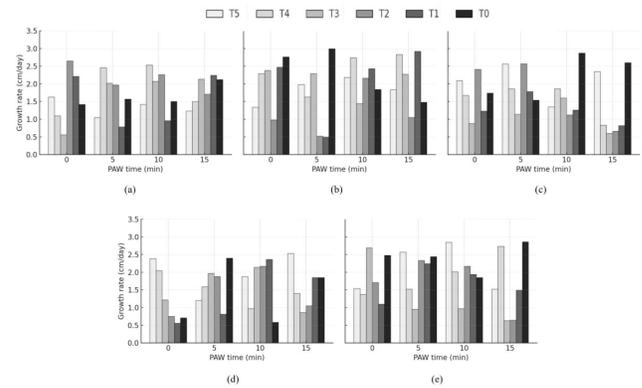

**Fig. 6 :** Growth rate of (a) WH 1142, (b) HI 1544, (c) GW 366, (d) GW 322, and (e) GW 273 respectively.

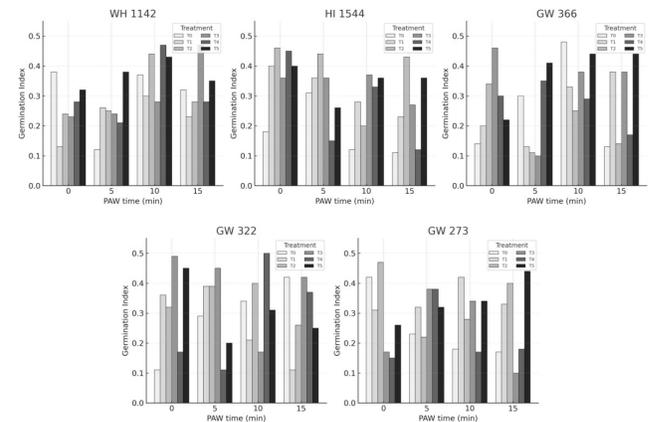

**Fig. 7 :** Germination index of (a) WH 1142, (b) HI 1544, (c) GW 366, (d) GW 322, and (e) GW 273 respectively.

Seedling growth responses varied across varieties, but were generally positive. The growth rate data in Fig. 6 demonstrate enhanced daily height increments for treated plants compared to controls. WH 1142 and GW 273 showed the most pronounced increases in root length under PAW irrigation, indicating that nitrate-rich PAW stimulated below-ground biomass. In contrast, GW 366 exhibited a strong shoot vigor, with elongated coleoptiles and faster leaf emergence, making it particularly suitable for late sowing under stress conditions. These trends suggest that plasma treatment not only improves average vigor, but also amplifies genetic predispositions of individual varieties.

Yield traits were also significantly influenced by plasma treatments. Spike length increased by 12–15% across varieties, as shown in Fig. 8, with the strongest response observed in GW 366. Grain number per spike improved markedly in GW 322, demonstrating that plasma exposure positively influenced reproductive development. Thousand-grain weight, a critical indicator of grain filling, improved in GW 273, confirming its enhanced assimilate partitioning and stress resilience. These yield improvements are supported by ANOVA results in Table - 2, which revealed significant differences between treatments at α = 0.05. The null hypothesis of no plasma effect was rejected, establishing plasma treatment as a statistically significant factor in improving wheat performance.

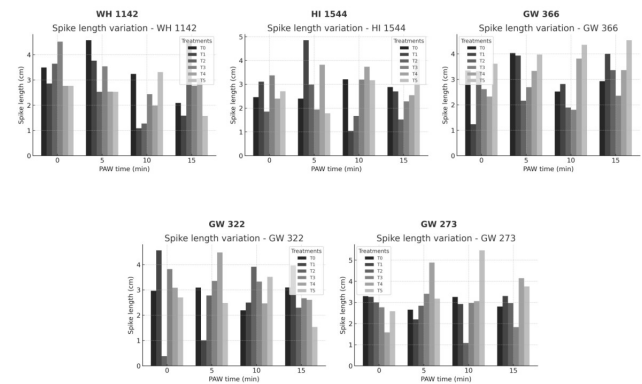

**Fig. 8 :** Spike length variation for (a) WH 1142, (b) HI 1544, (c) GW 366, (d) GW 322, and (e) GW 273 respectively.

| Trait | Source of Variation | F-value | p-value |
|---|---|---|---|
| Germination (%) | Treatment | 12.45 | <0.001 |
|  | Error | — | — |
| Root Length (cm) | Treatment | 8.32 | 0.002 |
|  | Error | — | — |
| Shoot Length (cm) | Treatment | 10.15 | <0.001 |
|  | Error | — | — |
| Yield (g/plant) | Treatment | 15.62 | <0.001 |
|  | Error | — | — |

*Significant at p < 0.05. ANOVA performed with treatment as fixed factor and replicates as random factor.*
**Table – 2:** ANOVA test results for growth, germination, and yield traits under different treatments

Variety-specific responses were noteworthy. WH 1142 exhibited superior early germination and uniform seedling establishment, consistent with its seed coat structure. HI 1544 responded most strongly in terms of yield, producing longer spikes and more grains per spike. GW 366 demonstrated the greatest improvement in grain number per spike, while GW 322 excelled in spike elongation. GW 273 developed the most robust root system and higher grain weight, indicating enhanced stress tolerance. These findings confirm that plasma effects are not uniform but depend on variety-specific physiological and morphological characteristics, emphasizing the need for tailored optimization under Indian agro-climatic conditions.

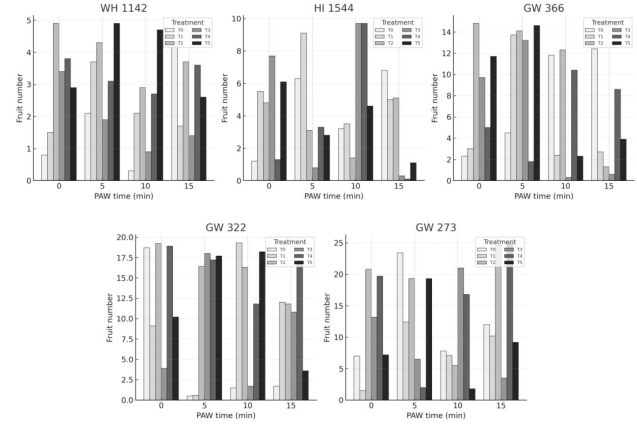

**Fig. 9 :** Fruit numbers for (a) WH 1142, (b) HI 1544, (c) GW 366, (d) GW 322, and (e) GW 273 respectively, on soaking.

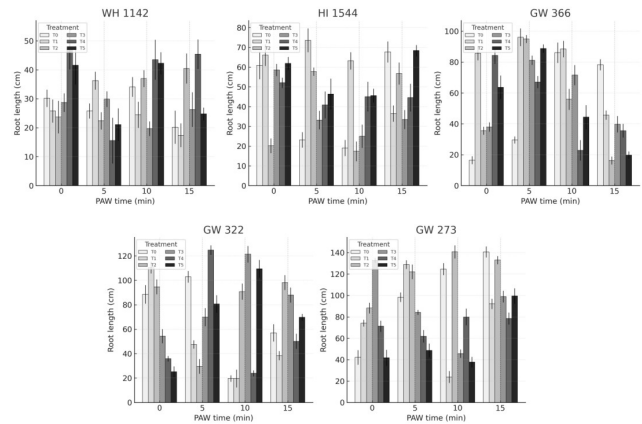

**Fig. 10 :** Root length of (a) WH 1142, (b) HI 1544, (c) GW 366, (d) GW 322, and (e) GW 273 respectively

Thus, plasma treatments significantly enhanced seed hydrophilicity, germination, seedling vigor, and yield traits in Indian wheat varieties. The observed effects are attributable to surface etching, increased water absorption, stimulation of metabolic activity by RONS, and nutritional benefits from PAW-derived nitrogen species. The results align with earlier reports on plasma-enhanced wheat germination and yield [17,18,27,28,35,36,37], but extend the findings by establishing variety-specific optimization for North Indian conditions. This study thus demonstrates that non-thermal plasma is a promising, chemical-free technology for improving wheat productivity, offering a sustainable solution for challenges of declining seed vigor and climate variability in Indian agriculture.

## IV. CONCLUSIONS

The present study demonstrates that non-thermal atmospheric pressure plasma, applied through dielectric barrier discharge (DBD) and gliding discharge, significantly improves wheat seed germination, seedling vigor, and yield. Comprehensive analyses of seed surface properties, germination potential, growth rate, spike length, fruit number, and root length reveal that plasma treatment increases seed wettability by reducing the water contact angle, with longer treatment times producing more pronounced effects. Direct plasma exposure as well as plasma-activated water (PAW) irrigation were both effective in enhancing germination characteristics, growth dynamics, and cultivation traits, indicating that plasma-generated reactive species, ultraviolet radiation, and OH radicals play a key role in modifying the seed surface and facilitating faster water uptake. Additionally, plasma treatment introduces active chemical species that penetrate the seed cell membrane, accelerating germination processes. Among the treatments tested, 3 minutes of direct DBD exposure and 15 minutes of PAW irrigation produced the most favorable outcomes, correlating with increased reactive nitrogen species (RNS), higher oxidation-reduction potential (ORP), and slightly reduced pH, which collectively contribute to improved germination and growth. These findings highlight atmospheric plasma treatment as a sustainable, environmentally friendly, and scalable technology with strong potential to enhance wheat productivity in North India. Future research should focus on field-scale trials across diverse agro-climatic zones, assessment under different wheat cultivars, and integration of plasma treatment protocols into national seed technology programs coordinated government and universities to fully realize its agronomic and economic benefits.